# Brookite $TiO_2$ as an active photocatalyst for photoconversion of plastic wastes to acetic acid and simultaneous hydrogen production: Comparison with anatase and rutile

Thanh Tam Nguyen[1,2], Kaveh Edalati[1,2,]*

[1] WPI, International Institute for Carbon Neutral Energy Research (WPI-I2CNER), Kyushu University, Fukuoka 819-0395, Japan

[2] Mitsui Chemicals, Inc. - Carbon Neutral Research Center (MCI-CNRC), Kyushu University, Fukuoka 819-0395, Japan

**Abstract**

Photoreforming is a clean photocatalytic technology for simultaneous plastic waste degradation and hydrogen fuel production, but there are still limited active and stable catalysts for this process. This work introduces the brookite polymorph of $TiO_2$ as an active photocatalyst for photoreforming with an activity higher than anatase and rutile polymorphs for both hydrogen production and plastic degradation. Commercial brookite successfully converts polyethylene terephthalate (PET) plastic to acetic acid under light. The high activity of brookite is attributed to good charge separation, slow decay and moderate electron trap energy, which lead to a higher generation of hydrogen and hydroxyl radicals and accordingly enhanced photo-oxidation of PET plastic. These results introduce brookite as a stable and active catalyst for the photoconversion of water contaminated with microplastics to value-added organic compounds and hydrogen.

*Corresponding author (E-mail: kaveh.edalati@kyudai.jp; Tel: +80-92-802-6744)

# 1. Introduction

Titanium oxide (TiO$_2$) exists in three natural crystalline polymorphs including anatase (tetragonal), brookite (orthorhombic) and rutile (tetragonal), as schematically shown in Fig. 1a [Lu et al., 2006; Murad, 1997; Haggerty et al., 2017]. In the crystal structure of anatase, each octahedron shares corners to generate (001) planes, while rutile is made up of edge-sharing octahedrons that create the (001) planes. The orthorhombic crystalline structure of brookite is composed of both corner- and edge-sharing octahedral crystals, making brookite crystal structure significantly less symmetric compared to rutile and anatase, [Chen et al., 2024]. While rutile is the thermodynamically stable phase of TiO$_2$, anatase and brookite are considered metastable phases which can remain stable under ambient conditions for prolonged periods. TiO$_2$ is renowned for its exceptional semiconducting, optical and catalytic properties, making it a highly efficient photocatalyst with superior stability [Ismael, 2019]. Other advantages of TiO$_2$ making it a common photocatalyst are low cost, nontoxicity, and chemical and thermal stability [Ismael, 2021]. However, the rapid charge recombination and broad bandgap energy limit its use in photocatalysis [Ismael, 2021]. TiO$_2$ has been used in many photocatalytic processes, including metal ions removal, organic chemical synthesis, organic compound remediation, CO$_2$ conversion and hydrogen production [Eddy et al., 2023; Lin and Lee, 2010; Huang et al., 2020, Chen et al., 2023; Cheng et al., 2017; Cheng et al., 2023; Ismael, 2020].

Characteristics of TiO$_2$ polymorphs, including crystal structure, surface features, particle morphology, crystallinity level, and amount of surface defects, all affect the photocatalytic activity [Bellardita et al., 2017]. Extensive research has explored the photocatalytic properties of TiO$_2$ polymorphs [Bacsa et al., 1998; Ohno et al., 2002; Yurdakal et al., 2008; Etacheri et al., 2011], with anatase emerging as the most extensively studied photocatalyst among all semiconductors due to its appropriate bandgap energy, greater kinetic stability, faster charge carrier separation and deeper charge carrier energy in the bulk [Linsebigler et al., 1995; Comparelli et al., 2005; Li et al., 2007; Porkodi and Arokiamary, 2007, Xu et al., 2011; Katal et al., 2020]. Rutile, characterized by the narrowest bandgap, generally exhibits lower photocatalytic activity than anatase [Odling and Robertson, 2015], although rutile demonstrates superiority for some reactions [Zhang et al., 2011; Djokić et al., 2020]. In contrast to anatase and rutile, relatively limited knowledge exists about the characteristics and photoactivity of brookite as a heterogenous photocatalyst due to the complications of its synthesis under normal manufacturing conditions [Di Paola et al., 2013].

However, brookite generally seems to have a high photocatalytic activity [Kominami et al., 2003; Li et al., 2004; Li et al., 2007; Zhang et al., 2011]. Brookite has a unique arrangement of octahedral $TiO_6$ in its crystal structure, which creates channels to form along the *c*-axis. Some smaller cations can bind to these channels, making these channels active sites for catalysis [Chen at al., 2024]. Furthermore, it was reported that oxygen atoms that are exposed on the {100} crystal plane of brookite can enhance the catalytic activity due to their unique connection to the octahedral chain [Chen at al., 2024]. The high activity of brookite has been reported for traditional photocatalytic reactions, such as production of hydrogen [Kominami et al., 2003], decomposition of toxic contaminants [Li et al., 2004; Li et al., 2007; Zhang et al., 2011; Ohtani et al., 1985; Zhao et al., 2009; Ismail t al., 2010], and conversion of $CO_2$ [Liu at al., 2012]. Despite these reports, the activity of brookite for some other applications including simultaneous photocatalytic plastic degradation and hydrogen production, which is also known as photoreforming, should be still investigated.

Photoreforming, a photocatalytic process for plastic degradation, is a green and effective technology that can harvest solar energy and utilize waste plastics to convert them into various useful products and simultaneously produce hydrogen [Tomoji and Tadayoshi, 1981]. As schematically shown in Fig. 1b, plastic waste acts as a sacrificial electron donor within this process, undergoing oxidation by reactive radicals and photogenerated holes in the valence band, while concurrently, photoexcited electrons engage in the reduction of water to generate hydrogen. Although $TiO_2$ was the first photocatalyst used for photoreforming in 1981 [Tomoji and Tadayoshi, 1981], there have been no attempts to examine the activity of its brookite phase for this process. To address the critical knowledge gap concerning the activity of brookite for photoreforming, a comparison of its activity with other $TiO_2$ polymorphs is imperative. Such a study is important because plastic and microplastic pollution is currently a global concern causing toxic effects on ecosystems and human beings [Guimarães et al., 2021; Rivers-Auty et al., 2023]. Moreover, hydrogen produced from the photocatalytic process is considered a green fuel that does not cause $CO_2$ emission [Martínez et al., 2020; Zeng et al., 2020].

In this study, for the first time, the brookite $TiO_2$ is utilized for photocatalytic plastic degradation and hydrogen production. It is shown that the activity of the brookite phase for the degradation of polyethylene terephthalate (PET) plastics and simultaneous hydrogen production is higher than anatase and rutile. These findings are discussed in terms of enhanced •OH radical formation on brookite due to its easy charge separation, slow decay and moderate electron trap depth.

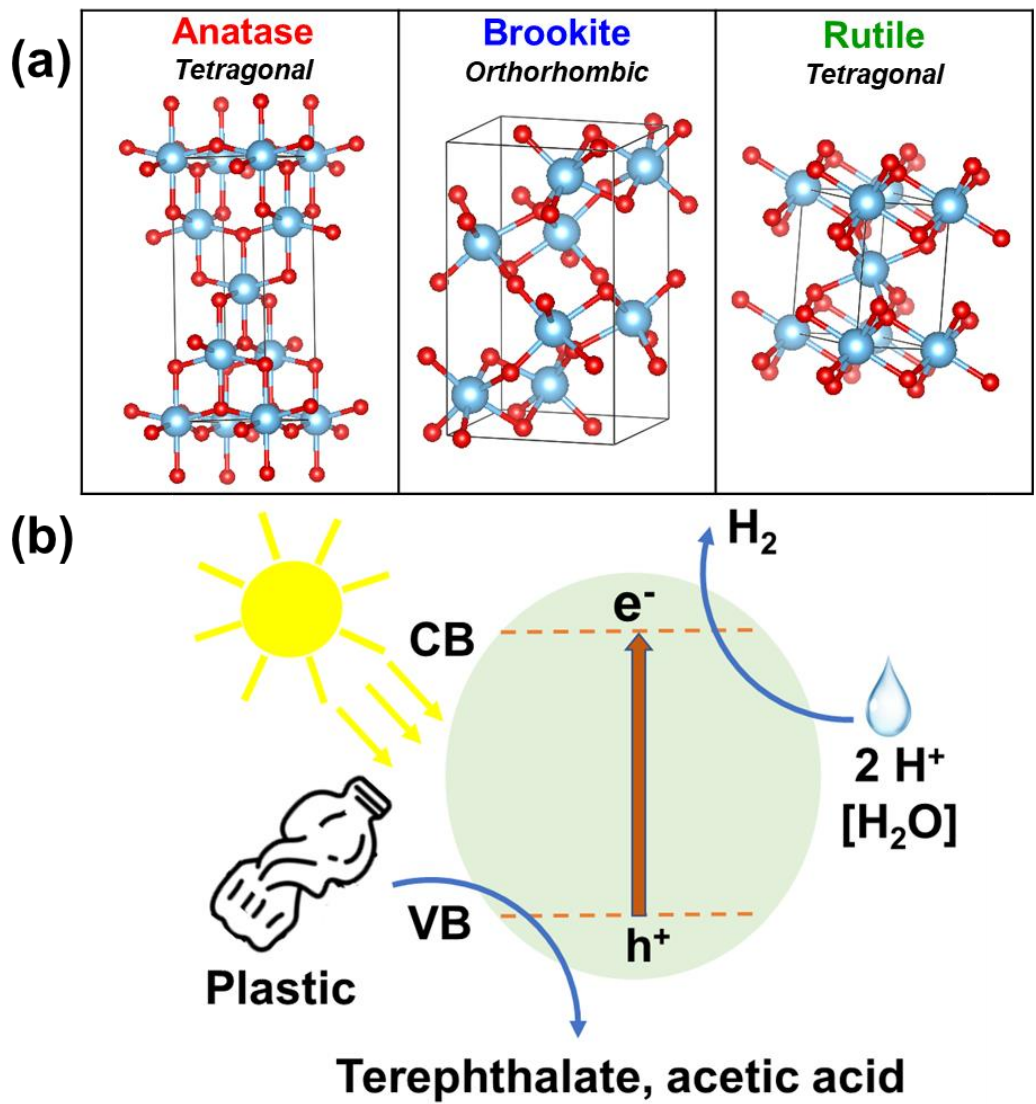

Figure 1. (a) Crystal lattice images of anatase, brookite and rutile. (b) Schematics of photoreforming for simultaneous photocatalytic plastic degradation and hydrogen production.

## 2. Materials and Methods
### 2.1. Reagents

Anatase (99.8%) powder was prepared from Sigma-Aldrich, USA, and brookite (99.99%) and rutile (99.99%) were purchased from Kojundo Chemical Company, Japan. NaOH was obtained from Fujifilm, Japan, and 10 M NaOD (40 wt% in D$_2$O with 99 at% purity) for nuclear magnetic resonance (NMR) tests was purchased from Sigma-Aldrich, Japan. PET plastic powder with an average particle size of 300 μm was obtained from GoodFellow, UK, to be used as a representative of microplastics in experiments. Coumarin was purchased from Tokyo Chemical Industry Co., LTD, Japan to examine the formation of hydroxyl (•OH) radicals.

## 2.2. Characterization of Catalysts

The crystalline structure of all catalysts including anatase, brookite, and rutile was characterized by X-ray diffraction (XRD) with Cu Kα irradiation and Raman spectroscopy with a 532 nm laser. The microstructure of catalysts was examined by scanning electron microscopy (SEM) under an acceleration voltage of 5 keV, and their crystallite size was estimated from the XRD profiles by the Halder-Wagner method [Halder and Wagner, 1962]. Light absorbance was measured by ultraviolet-visible light diffuse reflectance (UV-vis) spectroscopy and bandgap was estimated by the Kubelka-Munk analysis. The recombination of electrons and holes was investigated by the steady-state photoluminescence (PL) emission spectroscopy with a 325-nm laser source. The specific surface area was measured by nitrogen gas adsorption and the Brunauer-Emmett-Teller (BET) method.

## 2.3. Photocatalytic Experiments

For photocatalysis, 50 mg of catalyst and 50 mg of PET powder were mixed in 3 mL NaOH solution with a concentration of 10 M. After mixing, 250 µL of 0.01 M Pt(NH$_3$)$_4$(NO$_3$)$_2$ was added to the solution which is a source of platinum as a co-catalyst with 1 wt% of TiO$_2$ brookite, anatase or rutile. The mixture was then sonicated for 5 min, air-evacuated by argon for 30 min and irradiated by the full arc of a 300 W xenon lamp with an intensity of 18 kW/m$^2$ under continuous stirring and a constant temperature of 298 K. The produced hydrogen was measured by a gas chromatograph equipped with a thermal conductivity detector. The photo-oxidized products from PET plastics were identified by $^1$H NMR spectra using 10 M NaOD in D$_2$O. For quantitative analysis of $^1$H NMR, maleic acid was used as an internal standard. The level of hydroxyl (•OH) radicals formed during the photocatalytic process was evaluated by adding 1 mM of coumarin solution to the photocatalysis reactor. Since coumarin reacts with radicals through the following reaction, the concentration of 7-hydroxycoumarin was examined as an indication of (•OH) radicals [Ishibashi et al., 2000].

•OH + Coumarin → 7-Hydroxycoumarin  (1)

A spectrofluorometer was employed to record the fluorescence signal of the generated 7-hydroxycoumarin at 450 nm using an excitation wavelength of 332 nm. The cycling stability and reusability of the catalyst were further evaluated by three repeated photocatalytic cycles. After each cycle, the catalyst was collected, washed with deionized water until pH reached 7, dried at 333 K, and used for the photoreforming process again.

## 3. Results
### 3.1. Catalyst Characterization

Morphologies of samples examined by SEM are shown in Fig. 2. Anatase has round-shaped nanoparticles with ~130 nm average sizes. Brookite and rutile have larger particle sizes ranging from a few hundred nanometers to a few micrometers. The BET surface areas are 10.20, 0.55, and 0.71 m$^2$/g for anatase, brookite and rutile, respectively. These SEM and BET confirm the larger specific surface area for anatase. The XRD and Raman spectra of the three materials are shown in Fig. 3a and 3b, respectively. The crystal structure of each material is consistent with the reported tetragonal structure for anatase, orthorhombic structure for brookite and tetragonal structure for rutile [Lu et al., 2006; Murad, 1997; Haggerty et al., 2017]. The average crystallite size calculated by the Halder-Wagner method is 55, 28 and 85 nm for anatase, brookite and rutile, respectively. The ratios of particle size to crystal size indicate that there are a few crystals in each anatase particle, while brookite should have many nanocrystals in its particles. UV-vis diffuse reflectance spectra of the three catalysts are shown in Fig. 4. The three materials exhibit mainly light absorbance in the UV light region. The bandgap, estimated by the Kubelka-Munk theory, is 3.04 eV for anatase, 3.00 eV for brookite and 2.92 eV for rutile. These bandgap calculations are in good agreement with reported data by other researchers [Reyes-Coronado et al., 2008; Zhou et al., 2013; Zerjav et al., 2022]. The PL spectra of anatase, brookite, and rutile are shown in Fig. 5, implying that brookite expresses the lowest PL intensity among the investigated polymorphs, followed by rutile and anatase. This indicates that although the three polymorphs have similar light absorbance (i.e. similar electron-hole separation), brookite has the lowest irradiative recombination of generated charge carriers.

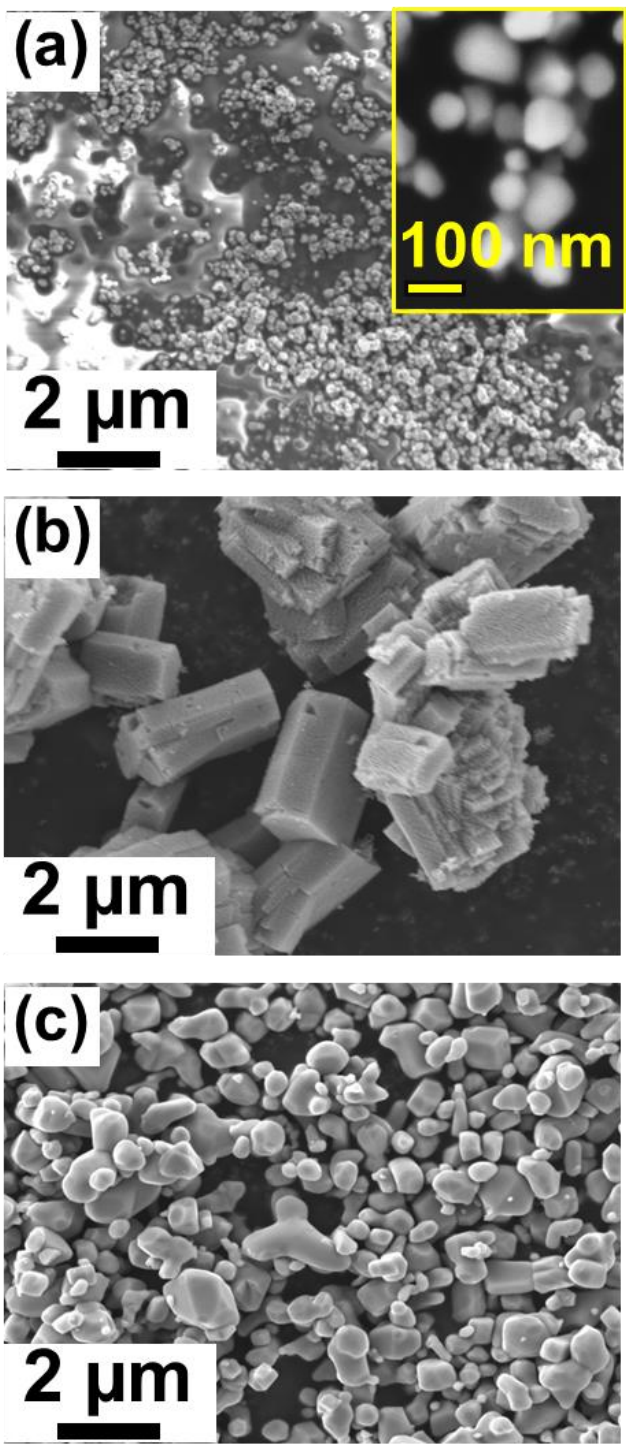

Figure 2. Morphologies of $TiO_2$ (a) anatase, (b) brookite and (c) rutile examined by SEM.

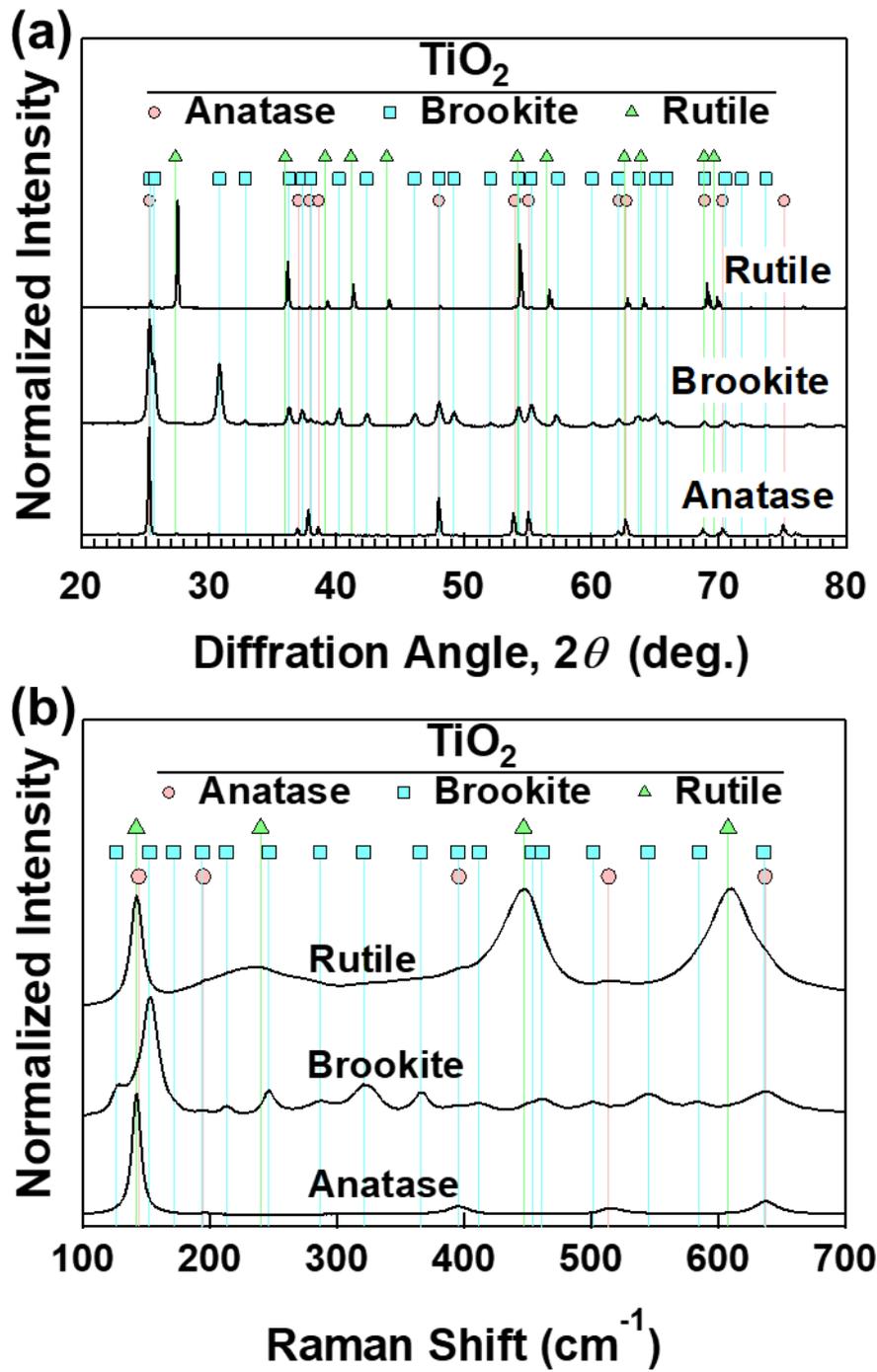

Figure 3. (a) XRD and (b) Raman spectra of anatase, brookite and rutile.

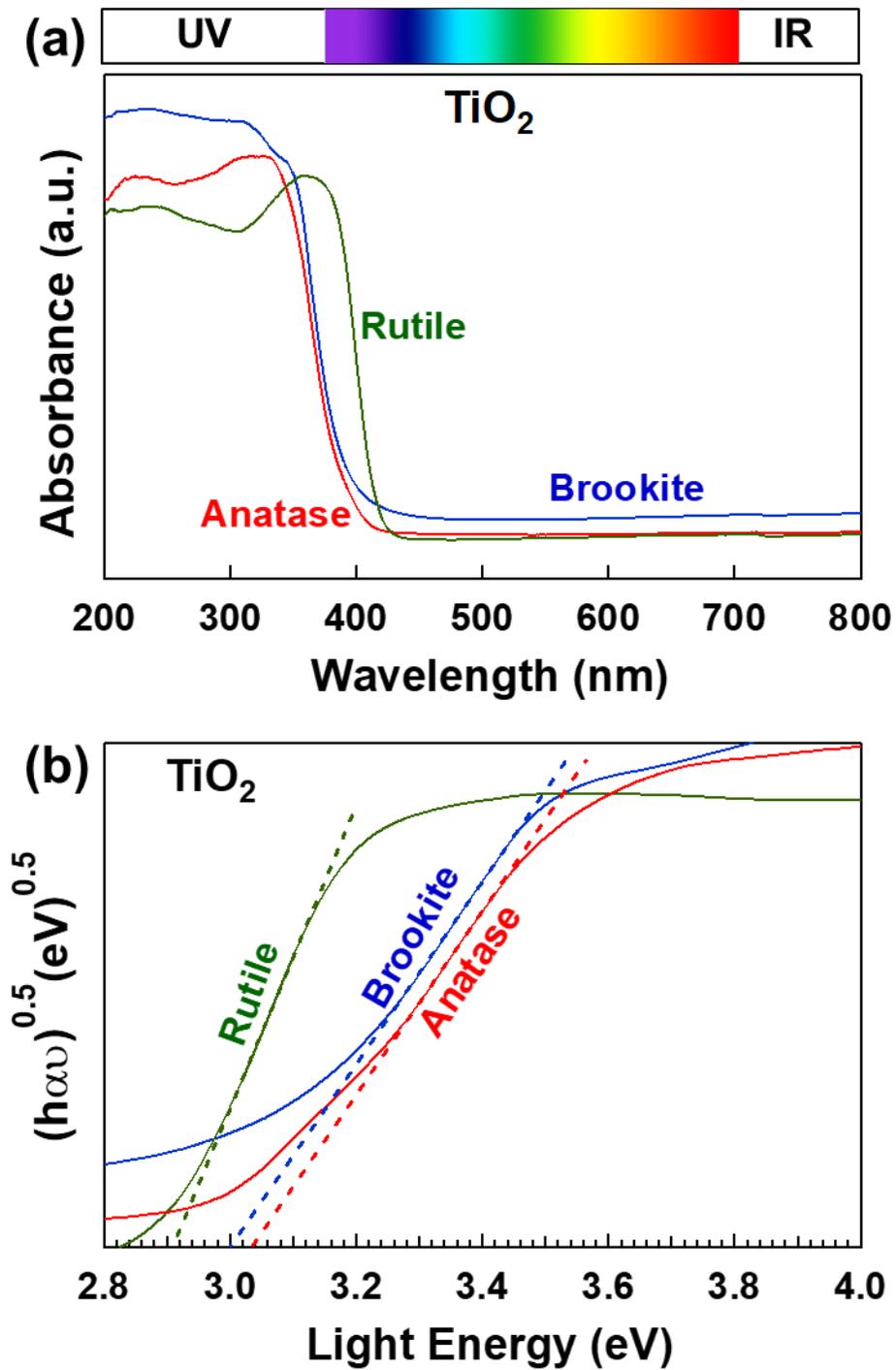

Figure 4. (a) UV-Vis light absorbance spectra and (b) corresponding Kubelka-Munk plot for bandgap calculation ($\alpha$: light absorption, h: Planck's constant, $\nu$: light frequency) for anatase, brookite and rutile.

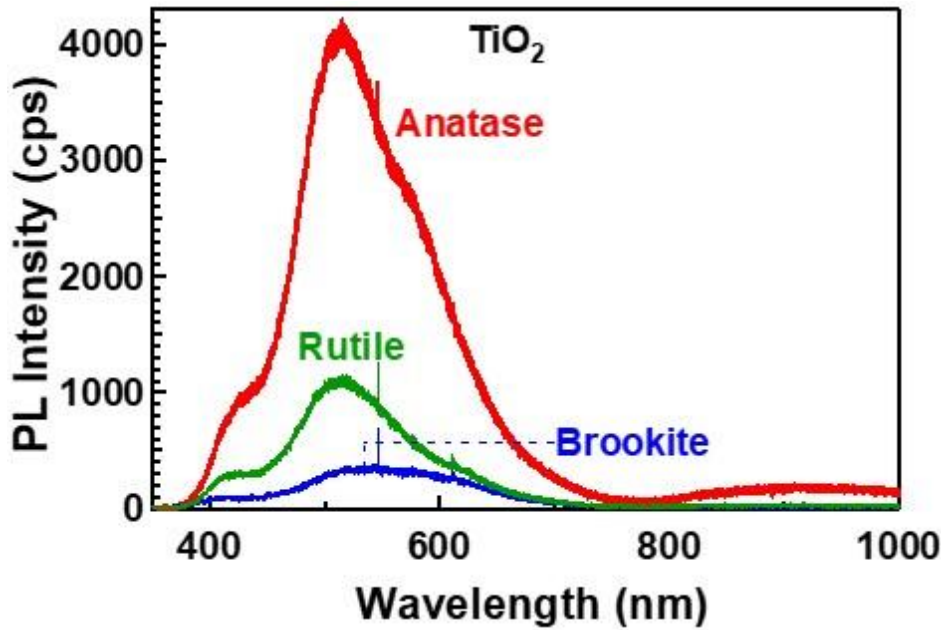

Figure 5. Photoluminescence spectrum using an excitation wavelength of 325 nm for anatase, brookite, and rutile.

## 3.2. Hydrogen Production

The photocatalytic activity of anatase, brookite, and rutile for hydrogen production during PET plastic degradation are shown in Fig. 6. No hydrogen is detected in the system without catalyst addition by light irradiation for 4 h. All the catalyst powders are active for photocatalytic hydrogen production with an activity in the order of anatase, rutile and brookite, but no oxygen is produced. After 4 h irradiation, the generated hydrogen amount is 12.38 mmol m$^{-2}$ for brookite which is about 28.5 and 1.7 times higher than that for anatase and rutile, respectively. Such a high activity of brookite for photoreforming is rather unexpected because anatase is usually considered the most active TiO$_2$ phase for conventional water splitting [Abe et al., 2001; Amano et al., 2009].

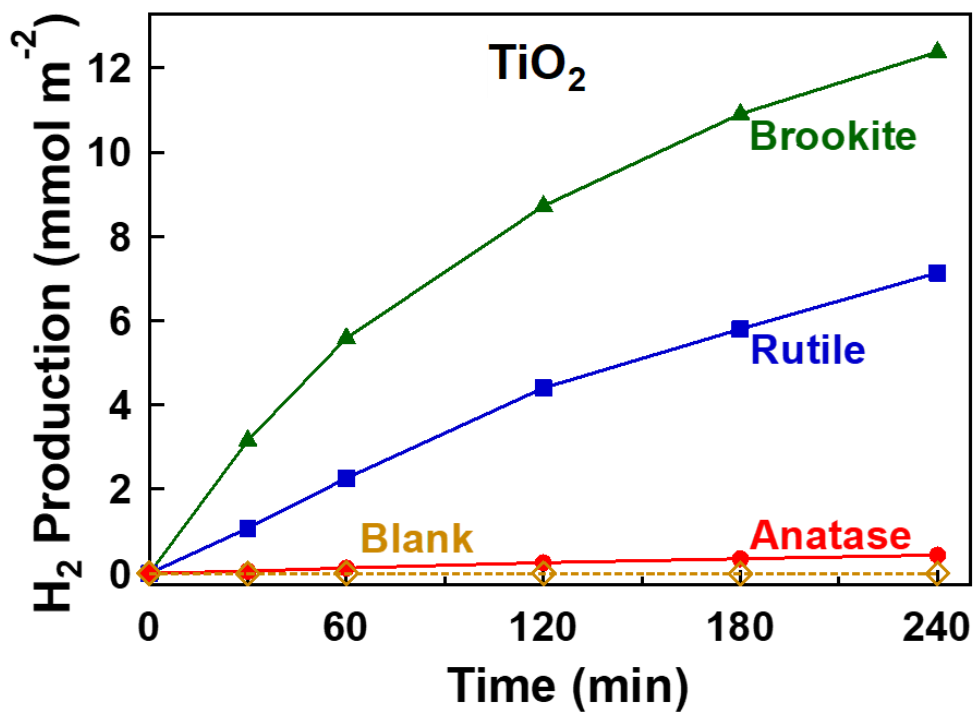

Figure 6. Photocatalytic hydrogen production from PET plastic degradation using anatase, brookite and rutile including blank test data.

## 3.3. PET Oxidation Products

The photoreforming process not only generates hydrogen in the reduction half-reaction as a useful product but also degrades plastic waste to valuable chemicals in the oxidation half-reaction. In this study, $^1$H NMR was used to identify the PET oxidation products after 4 h irradiation, as shown in Fig. 7. The $^1$H NMR spectrum of the blank test before the photocatalytic test in Fig. 7a shows only the peak of $D_2O$. All the $^1$H NMR spectra after photocatalysis using anatase, brookite, and rutile in Fig. 7b-d show the intense peak for $D_2O$, the standard peak of maleic acid and different new peaks for oxidation products. The photocatalytic process leads to the formation of terephthalate and acetic acid with different yields for different catalysts. The generation of terephthalate indicates that PET can be dissolved in the form of monomers in NaOH. As can be seen in Fig. 8, the amount of generated acetic acid increases in the order of anatase, rutile and brookite, indicating the great photocatalytic activity of brookite. The results are also in line with the high hydrogen production using brookite. Similar photo-degradation products for PET plastic were reported in previous studies using different photocatalysts such as $CdS/CdO_x$ quantum dots [Uekert et al., 2018], $CN_x/Ni_2P$ [Uekert et al., 2019] and CN-CNTs-NiMo hybrids [Gong et al., 2022], but the main difference is that brookite is chemically more stable and environmentally safer than previously used sulfides, phosphides and nitrides.

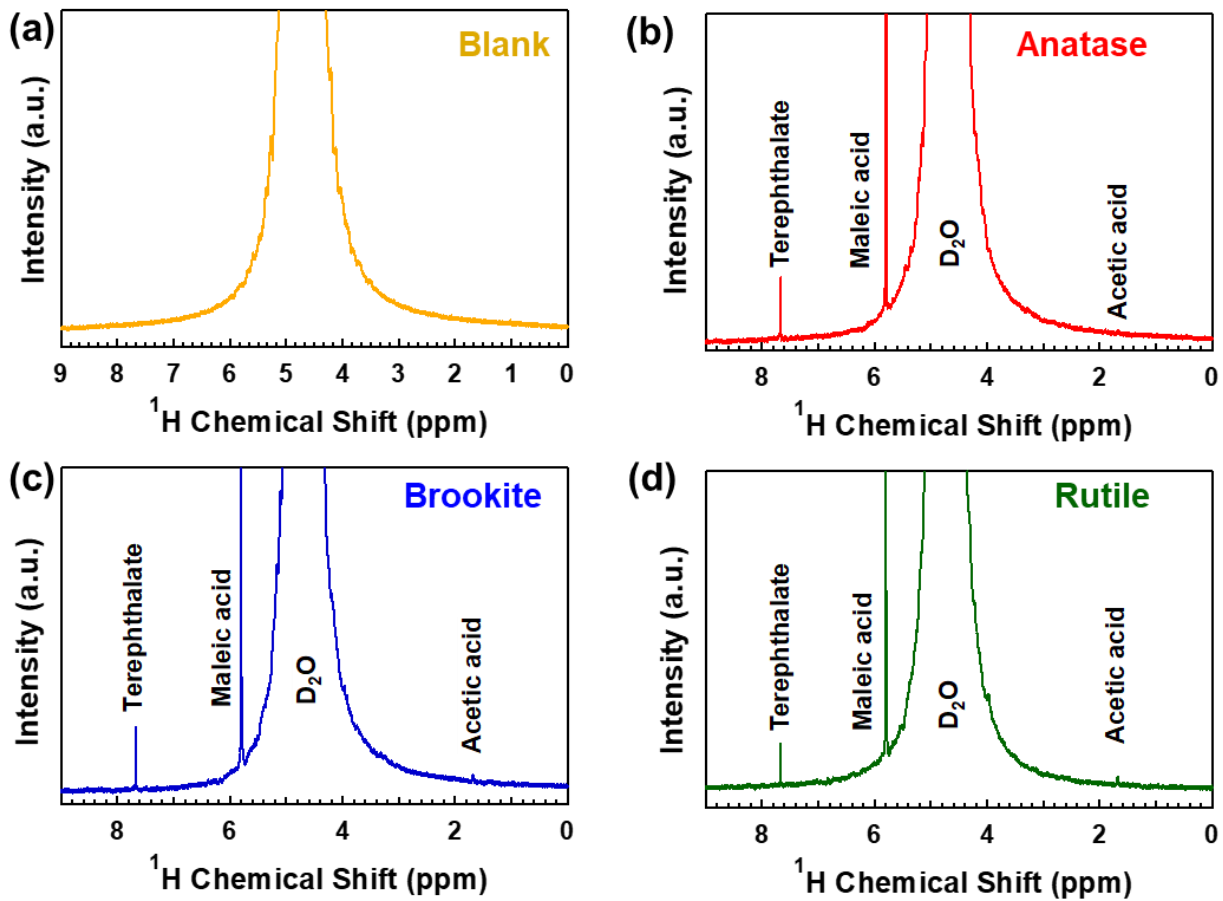

Figure 7. $^1$H NMR spectra of PET degradation products in 10 M NaOD in D$_2$O (a) before photocatalysis and after photocatalysis for 4 h using (b) anatase, (c) brookite and (d) rutile.

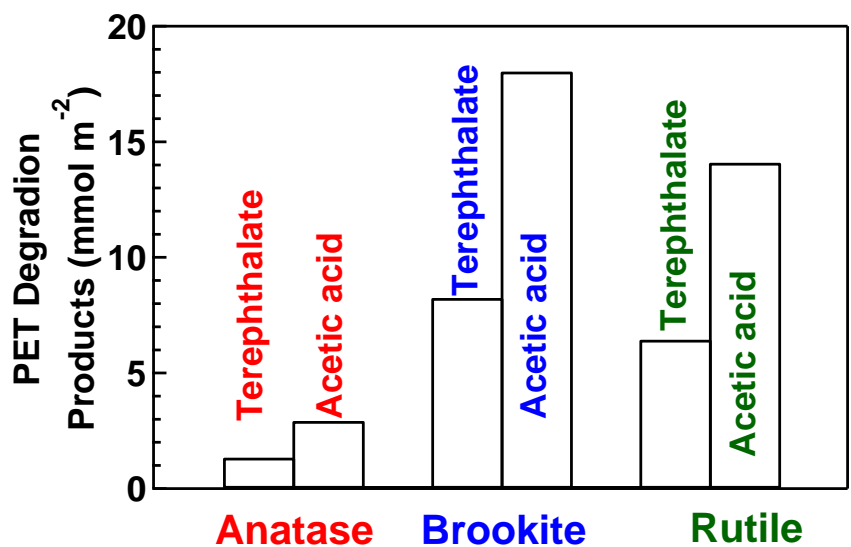

Figure 8. Photodegraded products from PET plastic. Quantitative $^1$H NMR analysis for concentration of terephthalate and acetic acid produced by degradation of PET after 4 h UV irradiation using anatase, brookite and rutile in solution of 10 M NaOD in D$_2$O with maleic acid as internal standard.

## 3.4. Hydroxyl Radical Generation

The formation of hydrogen gas and the absence of oxygen gas suggest that •OH radicals should have formed during photocatalysis. Moreover, since most photocatalytic degradation processes relate to oxidation transformation, as shown in the reactions below, it is crucial to assess the formation of •OH radicals below [Chu et al., 2022; Edirisooriya et al., 2023; Ashraf et al., 2023; Tang et al., 2023].

$Semiconductor + Photon \rightarrow h^+_{(Valence\ Band)} + e^-_{(Conduction\ Band)}$ (2)

$h^+ + H_2O \rightarrow •OH + H^+$ (3)

$2H^+ + 2e^- \rightarrow H_2$ (4)

•OH/holes + PET Plastics → Acetic Acid, Formic Acid, Glycolic Acid & Terephthalate (5)

In this study, the generation of •OH radicals during photocatalysis was evaluated using coumarin as a probe radical-trap chemical and the formation of 7-hydroxycoumarin as an indicator of •OH radical concentration was examined using fluorescence spectroscopy. Fig. 9 compares the fluorescence intensity measured after 4 h irradiation using anatase, brookite and rutile photocatalysts. The data show the increase of 7-hydroxycoumarin production in the order of anatase, rutile and brookite, indicating that brookite exhibits a higher formation rate of •OH radicals than other $TiO_2$ polymorphs. These results, which are in good agreement with photocatalytic hydrogen production results in Fig. 6 are in line with the reported high activity of brookite for other photocatalytic reactions [Zhang et al., 2011; Choi et al., 2017]. It should be noted not only •OH radicals but also holes can contribute to plastic degradation in photoreforming, as shown in Eq. (5) and reported in various studies [Chu et al., 2022; Edirisooriya et al., 2023; Ashraf et al., 2023; Tang et al., 2023]. The formation of •OH radicals is not limited to the photoreforming of plastics as it was reported in other photocatalytic processes for simultaneous hydrogen production and organic compound degradation [Jiang et al., 2018; Kang et al., 2023; Kumar et al., 2021].

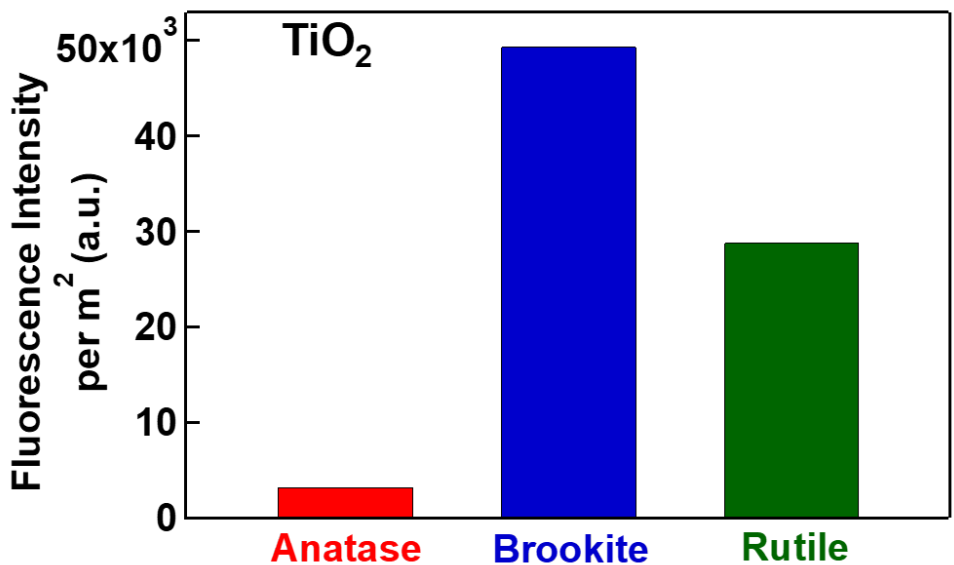

Figure 9. Fluorescence intensity of photocatalytic solution at a wavelength of 450 nm after 20 min irradiation in presence of coumarin as •OH radical trapping chemical and anatase, brookite and rutile as catalysts. Higher fluorescence intensity indicates higher concentration of •OH radicals.

### 3.5. Stability and Reusability of Photocatalysts

To study the stability of anatase, brookite, and rutile catalysts for simultaneous photocatalytic hydrogen production and PET plastic degradation, XRD and Raman measurements were conducted on catalysts before and after the photocatalytic reaction for 4 h. Raman gives information about a thin layer on the surface, while XRD provides information about the bulk of materials as well. As shown in Fig. 10, the crystal structures of all catalysts are stable as no changes in the XRD and Raman spectra are detected after photocatalysis.

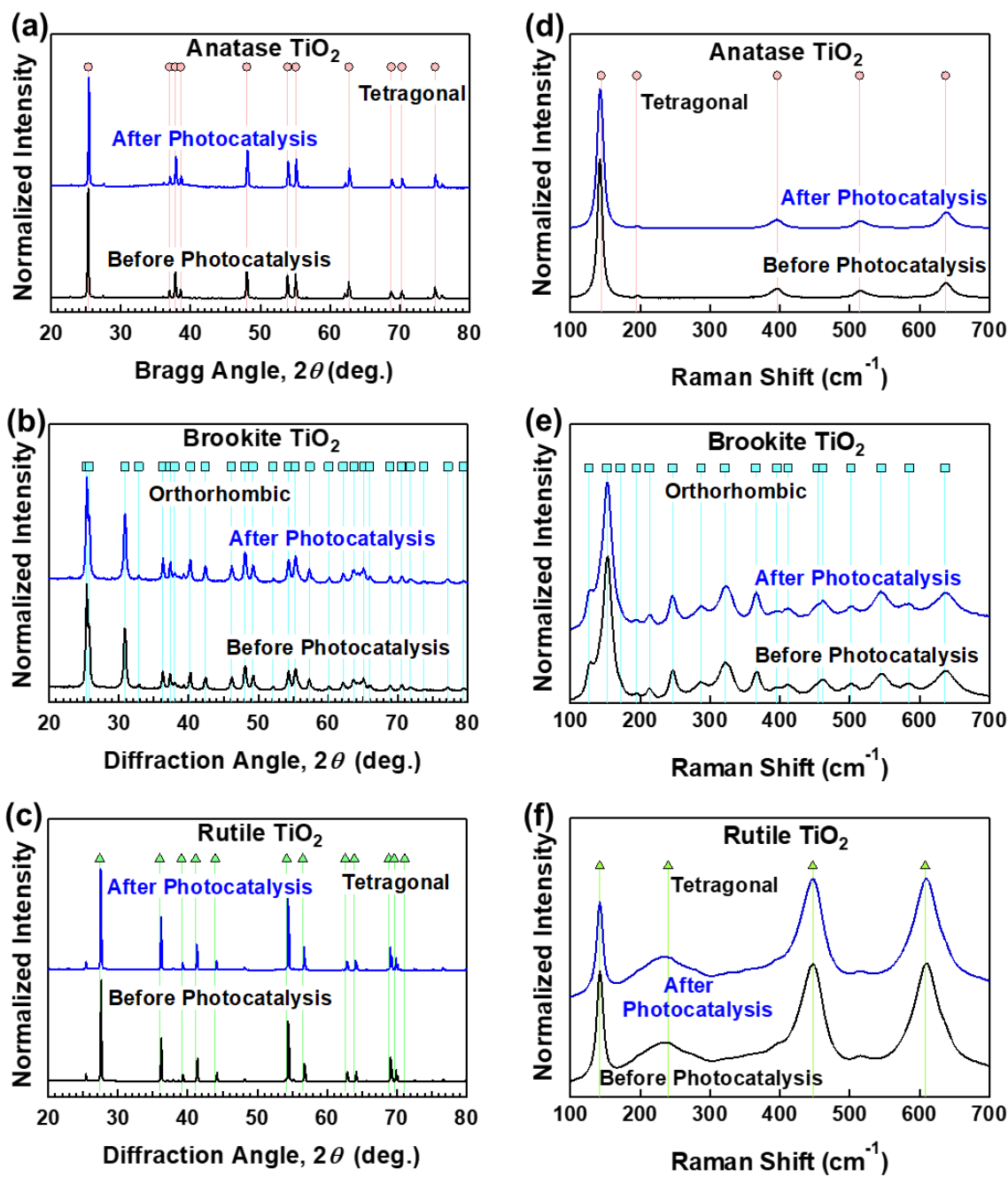

Figure 10. Stability of examined $TiO_2$ catalysts for simultaneous photocatalytic hydrogen production and PET plastic degradation. (a-c) XRD and (d-f) Raman spectra of (a, b) anatase, (c, d) brookite and (e, f) rutile before and after photocatalytic test for 4 h.

The cycling stability and reusability of the brookie catalyst in three repetitive photoreforming cycles are shown in Fig. 11. After three repeated photocatalytic tests, there is no significant change in the amount of hydrogen production, indicating that brookite has high stability and reusability for photocatalytic hydrogen production and plastic degradation. The high chemical stability of $TiO_2$ polymorphs compared to many other photocatalysts is one main reason that they have been widely used for various photocatalytic reactions [Odling and Robertson, 2015; Djokić et al., 2020; Zhang et al., 2011; Kominami et al., 2003; Zhang et al., 2012; Li et al., 2004; Zhao et al., 2012].

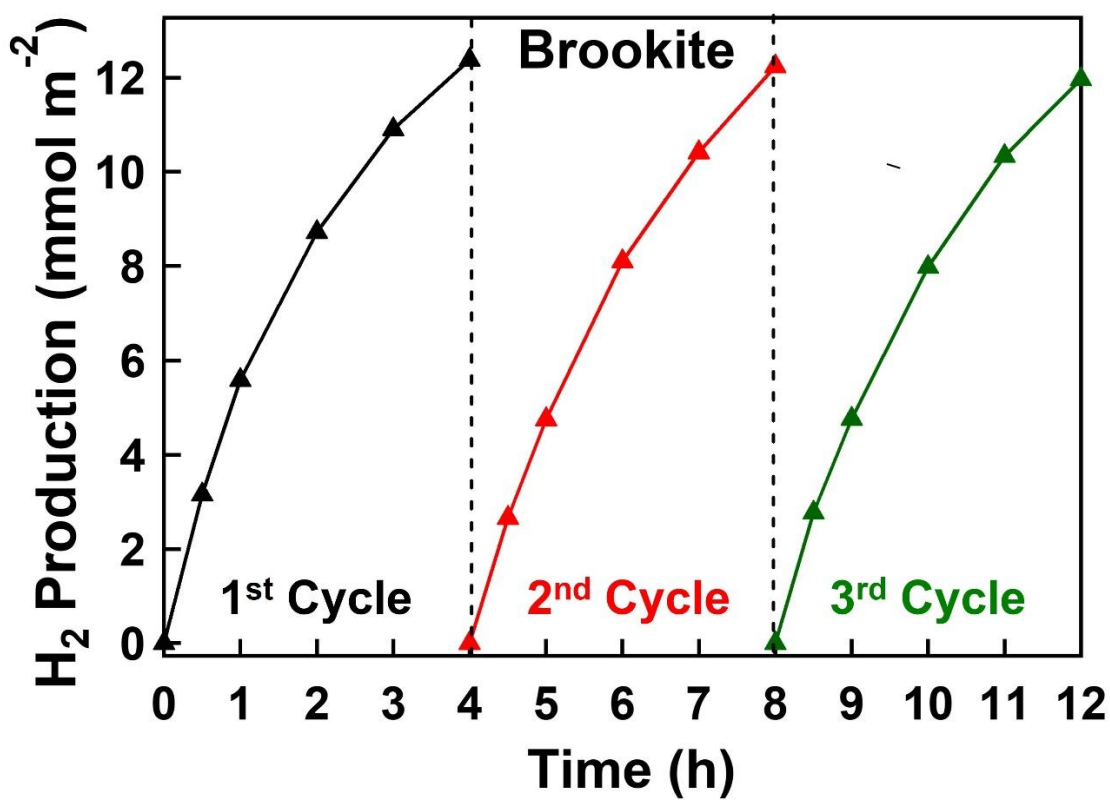

Figure 11. Reusability of brookite $TiO_2$ catalyst for simultaneous photocatalytic hydrogen production and PET plastic degradation.

## 4. Discussion

In this study, the photocatalytic activity of anatase, brookite, and rutile $TiO_2$ polymorphs for the photoreforming of plastic wastes and simultaneous hydrogen production is examined. The amounts of hydrogen and plastic degradation products are dependent on the polymorphism of TiO$_2$. It is found that the photocatalytic activity increases in the order of anatase, rutile and brookite. Since this is the first application of brookite for photoreforming, the reasons for its high activity need to be clarified. Two issues should be discussed in this regard in this section: (i) photocatalytic application of brookite for other reactions apart from photoreforming of plastic wastes, and (ii) mechanism of high photocatalytic activity of brookite by considering charge carrier separation, transport, recombination and trap depth.

The photocatalytic activity of brookite for other photocatalytic reactions was examined by different scientists, although the number of relevant publications is not high. Zhang *et al.* conducted experiments on photocatalytic degradation of Rhodamine B with anatase, brookite, and rutile [Zhang et al., 2011]. The Rhodamine B degradation efficiency increased in the order of anatase, rutile and brookite, which is consistent with the results of this study [Zhang et al., 2011]. Brookite was also reported to be a superior photocatalyst compared to anatase and rutile for the degradation and mineralization of some pharmaceutical compounds and phenol [Tran et al., 2017]. The photocatalytic activity of anatase, brookite, and rutile was also studied by comparing the degradation of tetramethylammonium and 4-chlorophenol [Choi et al., 2017]. Brookite revealed the highest photodegradation performance for both tetramethylammonium and 4-chlorophenol. The application of different TiO$_2$ polymorphs for photocatalytic hydrogen documented that the hydrogen production rates were 0.3719 μmol h$^{-1}$ m$^{-2}$ from anatase and 0.9742 μmol h$^{-1}$ m$^{-2}$ from brookite [Katai et al., 2024; Liu et al., 2014]. The high hydrogen production using brookite is consistent with this study, but the main difference is that the hydrogen production rate during photoreforming in Fig. 6 is ~5 times higher than the reported value for conventional water splitting. This indicates that photoreforming not only degrades plastic wastes and avoids using valuable sacrificial electron donors like methanol during water splitting but also leads to higher activity for hydrogen production.

Photocatalysis is still considered a complicated process that depends on many experimental factors and catalyst characteristics, and thus, it is challenging to clarify the exact mechanism underlying the high activity of a semiconductor like brookite. Based on the data gathered in this study and reported in the literature, the reasons for the higher photoreforming activity of brookite compared to other phases include the abundant •OH radical generation, the efficient separation of the photo-induced electron-hole pairs (i.e. easy separation and slow recombination), the charge carrier transport, and the moderate depth of electron-trap states. One of the main factors is the rate of charge carrier recombination. As shown in Fig. 5, the intensities of PL signals, representing the severity of irradiative recombination of photogenerated electrons and holes, increase in the order of brookite, rutile, and anatase. Therefore, brookite possesses the lowest irradiative recombination of photogenerated electron-hole pairs. Choi *et al.* examined the charge transfer kinetics of brookite and anatase and reported that the charge transfer resistance of brookite is significantly lower than that of anatase, making charge carrier separation easier in brookite [Choi et al., 2017]. Similar conclusions were achieved by utilizing open-circuit voltage decay which indicated that brookite possesses notably higher conductivity compared to anatase for facilitating the expeditious transport of photogenerated electrons [Choi and Yong, 2014]. Furthermore, time-resolved photoluminescence spectroscopy revealed that the decay process of brookite was slower than anatase. This indicates that charge carriers within brookite exhibit a slower recombination rate and a longer lifetime to produce hydrogen and •OH radicals as observed in current photoreforming experiments as well [Choi and Yong, 2014].

    Another reason attributed to the difference in photocatalytic activity of anatase, brookite, and rutile is the depth of charge carrier traps because shallow traps can enhance the reactivity of electrons on the one hand but decrease their lifetime on the other hand [Zerjav et al., 2022]. The moderate depth of trap states can ensure both electrons and holes remain reactive for their extended lifetime to contribute to the reduction and oxidation processes. In brookite, the presence of electron traps with moderate depth in the midgap states was identified, leading to the extension in both the quantity and lifetime of holes [Zerjav et al., 2022]. Examination of the behavior of photogenerated electrons and holes in $TiO_2$ by time-resolved visible to mid-infrared absorption spectroscopy also revealed that electron traps in brookite exhibit deeper depth than anatase and shallower depth than rutile [Vaquizo et al., 2017]. The study reported electron trap depths of 0.9, 0.4, and 0.1 eV for rutile, brookite, and anatase, respectively [Vaquizo et al., 2017]. It was found that the majority of photogenerated electrons in brookite are trapped at defects in a few picoseconds, leading to a reduced population of free electrons on the one hand and an extension of the lifetime of electrons on the other hand. The quantitative analyses by Vaquizo et al. confirmed that the number of holes increased in the order of anatase, rutile and brookite [Vequizo et al., 2017]. Since the large number of holes is a favorable parameter for enhancing photocatalytic oxidation reactions, these results provide another explanation for the high activity of brookite for photoreforming of plastic wastes.

In summary, the preeminent photocatalytic activity of brookite compared to rutile and anatase can be attributed to a combination of factors, including efficient charge separation, slow decay process, moderate charge carrier trap depth and robust •OH radical generation. Although the use of $TiO_2$ for photoreforming has been virtually abandoned due to the reported low activity of anatase, the current results open a pathway to use brookite $TiO_2$ for this application. The use of brookite provides some other advantages such as high stability and environmentally friendly features. Significant dependence of photoreforming activity on crystal structure suggests that other polymorphs of $TiO_2$ such as the high-pressure columbite phase, which shows high activity for hydrogen production [Razavi-Khosroshahi et al., 2016; Hidalgo-Jiménez et al., 2023] and $CO_2$ conversion [Akrami et al., 2021], and mixture of different $TiO_2$ phases such as P25 catalyst [Janus and Morawski, 2007; Chen et al., 2021] deserves to be examined for photoreforming in future studies.

## 5. Conclusions

In this work, anatase, brookite and rutile photocatalysts were used for the photoreforming process, which included hydrogen production and simultaneous PET plastic degradation to terephthalate and acetic acid. A thorough comparison of the properties of catalysts and their photocatalytic performance showed clear distinctions, with brookite emerging as the most active photocatalyst, followed by rutile and anatase. The high activity of brookite, which confirmed the importance of the crystal structure on photoreforming, was attributed to its efficient charge transfer kinetics, slow decay rate, abundant •OH radical generation, and moderate depth of electron-trap states.

## CRediT Authorship Contribution Statement

T. T. Nguyen and K. Edalati: Conceptualization, Methodology, Validation, Investigation, Writing – original draft, and Writing – Review & Editing.

## Declaration of Competing Interest

The authors declare that they have no known competing financial interests or personal relationships that could have appeared to influence the work reported in this paper.

## Data Availability

The data that support the findings of this study are available from the corresponding author upon request.

## Acknowledgments

This study is supported partly by Mitsui Chemicals, Inc., Japan, and partly through Grants-in-Aid from the Japan Society for the Promotion of (JP19H05176 & JP21H00150).

## References

Abe, R., Sayama, K., Domen, K., & Arakawa, H., 2001. A new type of water splitting system composed of two different $TiO_2$ photocatalysts (anatase, rutile) and a $IO_3^-/I^-$ shuttle redox mediator. Chem. Phys. Lett. 344, 339–344.

Akrami, S., Watanabe, M., Ling, T. H., Ishihara, T., Arita, M., Fuji, M., Edalati, K., 2021. High-pressure $TiO_2$-II polymorph as an active photocatalyst for $CO_2$ to CO conversion. Appl. Catal. B 298, 120566.

Amano, F., Prieto-Mahaney, O.-O., Terada, Y., Yasumoto, T., Shibayama, T., Ohtani, B., 2009. Decahedral single-crystalline particles of anatase titanium (IV) oxide with high photocatalytic activity, Chem. Mater. 21, 2601–2603.

Ashraf, M., Ullah, N., Khan, I., Tremel, W., Ahmad, S., Tahir, M. N., 2023. Photoreforming of waste polymers for sustainable hydrogen fuel and chemicals feedstock: waste to energy. Chem. Rev. 123, 4443–4509.

Bacsa, R. R., Kiwi, J., 1998. Effect of rutile phase on the photocatalytic properties of nanocrystalline titania during the degradation of p-coumaric acid. Appl. Catal. B 16, 19–29.

Bellardita, M., Di Paola, A., Megna, B., Palmisano, L., 2017. Absolute crystallinity and photocatalytic activity of brookite $TiO_2$ samples. Appl. Catal. B 201, 150–158.

Chen, S., Qian, T. T., Zhu, R., Jiang, H., 2021. Integrating modeling and experimental method for narrowing the optimum phase composition in P25 photocatalyst for typical aromatic pollutants degradation, Chem. Eng. J. 417, 128061.

Chen, Z., Tong, X., Cheng, G., 2023. A comparative study on metal species implanted amine-brookite-$TiO_2$ nanorods for enhanced photocatalytic $CO_2$ reduction, Carbon Lett. 33, 2041-2051.

Chen, Z., Xiong, J., Cheng, G., 2024. Recent advances in brookite phase TiO$_2$-based photocatalysts toward CO$_2$ reduction, Fuel 357, 129806.

Chen, W., Xiong, J., Wen, Z., Chen, R., Cheng, G., 2023. Synchronistic embedding of oxygen vacancy and Ag nanoparticles into potholed TiO$_2$ nanoparticles-assembly for collaboratively promoting photocatalytic CO$_2$ reduction, Mol. Catal. 542, 113138.

Cheng, G., Wei, Y., Xiong, J., Gan, Y., Zhu, J., Xu, F., 2017. Same titanium glycolate precursor but different products: successful synthesis of twinned anatase TiO$_2$ nanocrystals with excellent solar photocatalytic hydrogen evolution capability, Inorg. Chem. Front. 4, 1319.

Choi, M, Lim, J., Baek, M., Choi, W., Kim, W., Yong, K., 2017. Investigating the unrevealed photocatalytic activity and stability of nanostructured brookite TiO$_2$ film as an environmental photocatalyst, ACS Appl. Mater. Interfaces 9, 16252–16260.

Choi, M., Yong, K., 2014. A facile strategy to fabricate high-quality single crystalline brookite TiO$_2$ nanoarrays and their photoelectrochemical properties. Nanoscale 6, 13900–13909.

Chu, S., Zhang, B., Zhao, X., Soo, H. S., Wang, F., Xiao, R., Zhang, H., 2022. Photocatalytic conversion of plastic waste: from photodegradation to photosynthesis. Adv. Energy Mater. 12, 202200435.

Comparelli, R., Fanizza, E., Curri, M. L., Cozzoli, P. D., Mascolo, G., Passino, R., Agostiano, A., 2005. Photocatalytic degradation of azo dyes by organic-capped anatase TiO$_2$ nanocrystals immobilized onto substrates. Appl. Catal. B. 55, 81–91.

Di Paola, A., Bellardia, M., Palmisano, L., 2013. Brookite, the least known TiO$_2$ photocatalyst. Catalysts 3, 36–73.

Djokić, V. R., Marinković, A. D., Petrović, R. D., Ersen, O., Zafeiratos, S., Mitrić, M., Ophus, C., Radmilović, V. R., Janaćković, D. T., 2020. Highly active rutile TiO$_2$ nanocrystalline photocatalysts. ACS Appl. Mater. Interfaces 12, 33058–33068.

Eddy, D. R., Permana, M. D., Sakti, L. K., Sheha, G. A. N., Solihudin, Hidayat, S., Takei, T., Kumada, N., Rahayu, I., 2023. Heterophase polymorph of TiO$_2$ (anatase, rutile, brookite, TiO$_2$ (B)) for efficient photocatalyst: fabrication and activity. Nanomaterials 13,13040704.

Edirisooriya, E. M. N. T., Senanayake, P. S., Wang, H. B., Talipov, M. R., Xu, P., Wang, H., 2023. Photo-reforming and degradation of waste plastics under UV and visible light for H$_2$ production using nanocomposite photocatalysts, J. Environ. Chem. Eng. 11, 109580.

Etacheri, V., Seery, M. K., Hinder, S. J., Pillai, S. C., 2011. Oxygen rich titania: A dopant free, high temperature stable, and visible-light active anatase photocatalyst. Adv. Funct. Mater. 21, 3744–3752.

Gong, X., Tong, F., Ma, F., Zhang, Y., Zhou, P., Wang, Z., Liu, Y., Wang, P., Cheng, H., Dai, Y., Zheng, Z., Huang, B., 2022. Photoreforming of plastic waste poly (ethylene terephthalate) via in-situ derived CN-CNTs-NiMo hybrids. Appl. Catal. B 307, 121143.

Guimarães, A. T. B., Charlie-Silva, I., Malafaia, G., 2021. Toxic effects of naturally-aged microplastics on zebrafish juveniles: a more realistic approach to plastic pollution in freshwater ecosystems. J. Hazard. Mater. 407, 124833.

Haggerty, J. E. S., Schelhas, L. T., Kitchaev, D. A., Mangum, J. S., Garten, L. M., Sun, W., Stone, K. H., Perkins, J. D., Toney, M. F., Ceder, G., Ginley, D. S., Gorman, B. P., Tate, J., 2017. High-fraction brookite films from amorphous precursors. Sci Rep 7, 15232.

Halder, R.P.I., Wagner, C.N.J., 1962. X-Ray diffraction study of the effects of solutes on the occurrence of stacking faults in silver-base alloys, J. Appl. Phys. 33, 3451–3458.

Hidalgo-Jiménez, J., Akbay, T., Ishihara, T., Edalati, K., 2023. Understanding high photocatalytic activity of the $TiO_2$ high-pressure columbite phase by experiments and first-principles calculations. J. Mater. Chem. A 11, 23523–23535.

Huang, J., Li, D., Li, R., Chen, P., Zhang, Q., Liu, H., Lv, W., Liu, G., Feng, Y., 2020. One-step synthesis of phosphorus/oxygen co-doped $g-C_3N_4$/anatase $TiO_2$ Z-scheme photocatalyst for significantly enhanced visible-light photocatalysis degradation of enrofloxacin. J. Hazard. Mater. 386, 121634.

Ishibashi, K., Fujishima, A., Watanabe, T., Hashimoto, K., 2000. Detection of active oxidation species in $TiO_2$ photocatalysis using the fluorescence technique. Electrochem. Commun. 2, 207-210.

Ismael, M., 2019. Highly effective ruthenium-doped $TiO_2$ nanoparticles photocatalyst for visible-light-driven photocatalytic hydrogen production. New J. Chem. 43, 9596.

Ismael, M., 2020. Enhanced photocatalytic hydrogen production and degradation of organic pollutants from Fe (III) doped $TiO_2$ nanoparticles. J. Environ. Chem. Eng. 8, 103676.

Ismael, M., 2021. Ferrites as solar photocatalytic materials and their activities in solar energy conversion and environmental protection: A review. Sol. Energy Mater Sol. Cells 219, 110786.

Ismael, M., 2021. Latest progress on the key operating parameters affecting the photocatalytic activity of TiO$_2$-based photocatalysts for hydrogen fuel production: a comprehensive review. Fuel 303, 121207

Ismail, A. A., Kandiel, T. A., Bahnemann, D. W., 2010. Novel (and better?) titania-based photocatalysts: Brookite nanorods and mesoporous structures. J. Photochem. Photobiol. A Chem. 216, 183–193.

Janus, M., Morawski, A. W., 2007. New method of improving photocatalytic activity of commercial Degussa P25 for azo dyes decomposition. Appl. Catal. B 75, 118–123.

Jiang, X-H., Wang, L-C., Yu, F., Nie, Y-C., Xing, Q-J., Liu, X., Pei,.Y., Zou, J-P., Dai, W-L., 2018. Photodegradation of organic pollutants coupled with simultaneous photocatalytc evolution of hydrogen using quantum-dot-modified g-C$_3$N$_4$ catalysts under visible-light irradiation, ACS Sustainable Chem. Eng. 6, 12695–12705.

Kang, F., Shi, C., Zhu, Y., Eqi, M., Shi, J., Teng, M., Huang, Z., Si, C., Jiang, F., Hu, J., 2023. Dual-functional marigold-like Zn$_x$Cd$_{1-x}$S homojunction for selective glucose photoreforming with remarkable H$_2$ coproduction. J. Energy Chem. 79, 158–167.

Katai, M., Edalati, P., Hidalgo-Jimenez, J., Shundo, Y., Akbay, T., Ishihara, T., Arita, M., Fuji, M., Edalati, K., 2024. Black brookite rich in oxygen vacancies as an active photocatalyst for CO$_2$ conversion: experiments and first-principles calculations. J. Photochem. Photobiol. A 449, 115409.

Katal, R., Masudy-Panah, S., Tanhaei, M., Farahani, M. H. D. A., Jiangyong, H., 2020. A review on the synthesis of the various types of anatase TiO$_2$ facets and their applications for photocatalysis. J. Chem. Eng. 384, 123384.

Kominami, H., Ishii, Y., Kohno, M., Konishi, S., Kera, Y., Ohtani, B., 2003. Nanocrystalline brookite-type titanium (IV) oxide photocatalysts prepared by a solvothermal method: correlation between their physical properties and photocatalytic activities. Catal. Letters 91, 41–47.

Kumar, A., Sharma, G., Kumari, A., Guo, C., Naushad, M., Vo, DVN., Iqbal, J., Stadler, FJ., 2021. Construction of dual Z-scheme g-C$_3$N$_4$/Bi$_4$Ti$_3$O$_{12}$/Bi$_4$O$_5$I$_2$ heterojunction for visible and solar powered coupled photocatalytic antibiotic degradation and hydrogen production: Boosting via I-/I3- and Bi3+/Bi5+ redox mediators. Appl. Catal. B 284, 119808.

Li, J. G., Ishigaki, T., Sun, X., 2007. Anatase, brookite, and rutile nanocrystals via redox reactions under mild hydrothermal conditions: Phase-selective synthesis and physicochemical properties. J. Phys. Chem. C 111, 4969–4976.

Li, J. G., Ishigaki, T., Sun, X., 2007. Anatase, brookite, and rutile nanocrystals via redox reactions under mild hydrothermal conditions: Phase-selective synthesis and physicochemical properties. J. Phys. Chem. C 111, 4969–4976.

Li, J. G., Tang, C., Li, D., Haneda, H., Ishigaki, T., 2004. Monodispersed spherical particles of brookite-type $TiO_2$: Synthesis, characterization, and photocatalytic property. J. Am. Ceram. 87, 1358–1361.

Lin, Y. C., Lee, H. S., 2010. Effects of $TiO_2$ coating dosage and operational parameters on a $TiO_2$/Ag photocatalysis system for decolorizing Procion red MX-5B. J. Hazard. Mater. 179, 462-470.

Linsebigler, A. L., Lu, G., Yates, J. T., 1995. Photocatalysis on $TiO_2$ Surfaces: Principles, Mechanisms, and Selected Results. Chem. Rev. 95, 735–758.

Liu, L., Zhao, H., Andino, J. M., Li, Y., 2012. Photocatalytic $CO_2$ reduction with $H_2O$ on $TiO_2$ nanocrystals: Comparison of anatase, rutile, and brookite polymorphs and exploration of surface chemistry. ACS Catal. 2, 1817–1828.

Liu, Y., Wang, Z., Wang, W., Huang, W., 2014. Engineering highly active $TiO_2$ photocatalysts via the surface-phase junction strategy employing a titanate nanotube precursor. J. Catal. 310, 16–23.

Lu, Y., Jaeckel, B., Parkinson, B. A., 2006. Preparation and characterization of terraced surfaces of low-index faces of anatase, rutile, and brookite. Langmuir 22, 4472–4475.

Martínez, M. C. N., Bajorowicz, B., Klimczuk, T., Żak, A., Łuczak, J., Lisowski, W., & Zaleska-Medynska, A. (2020). Synergy between $AgInS_2$ quantum dots and ZnO nanopyramids for photocatalytic hydrogen evolution and phenol degradation. J. Hazard. Mater. 398, 123250.

Murad, E., 1997. Identification of minor amounts of anatase in kaolins by Raman spectroscopy. Am. Mineral. 82, 203–206.

Odling, G., Robertson, N., 2015. Why is anatase a better photocatalyst than rutile? the importance of free hydroxyl radicals. Chem. Sus. Chem. 8, 1838–1840.

Ohno, T., Sarukawa, K., Matsumura, M., 2002. Crystal faces of rutile and anatase $TiO_2$ particles and their roles in photocatalytic reactions. New J. Chem. 26, 1167–1170, 2002.

Ohtani, B., Handa, J., Nishimoto, S., Kagiya, T., 1985. Highly active semiconductor photocatalyst: Extra-fine crystallite of brookite TiO$_2$ for redox reaction in aqueous propan-2-ol and/ or silver sulfate solution. Chem. Phys. Lett. 120, 292–294.

Porkodi, K., Arokiamary, S. D., 2007. Synthesis and spectroscopic characterization of nanostructured anatase titania: A photocatalyst. Mater. Charact. 58, 495–503.

Razavi-Khosroshahi, H., Edalati, K., Hirayama, M., Emami, H., Arita, M., Yamauchi, M., Hagiwara, H., Ida, S., Ishihara, T., Akiba, E., Horita, Z., Fuji, M., 2016. Visible-light-driven photocatalytic hydrogen generation on TiO$_2$-II stabilized by high-pressure torsion. ACS Catal. 6, 5103–5107.

Reyes-Coronado, D., Rodríguez-Gattorno, G., Espinosa-Pesqueira, M. E., Cab, C., De Coss, R., Oskam, G., 2008. Phase-pure TiO$_2$ nanoparticles: Anatase, brookite and rutile. Nanotechnology 19, 145605.

Rivers-Auty, J., Bond, A. L., Grant, M. L., & Lavers, J. L., 2023. The one-two punch of plastic exposure: macro-and micro-plastics induce multi-organ damage in seabirds. J. Hazard. Mater. 442, 130117.

Tang, X., Han, X., Sulaiman, N. H. M., He, L., Zhou, X., 2023. Recent advances in the photoreforming of plastic waste: principles, challenges, and perspectives. Ind. Eng. Chem. Res. 62, 9032–9045.

Tomoji, K., and Tadayoshi, S., 1981. Photocatalytic hydrogen production from water by the decomposition of poly-vinylchroride, protein, algae, dead insects, and excrement. Chem. Lett. 10, 81-84.

Tran, H. T. T., Kosslick, H., Ibad, M. F., Fischer, C., Bendtrup, U., Vuong, T. H., Nguyen, L. Q., Schulz, A., 2017. Photocatalytic performance of highly active brookite in the degradation of hazardous organic compounds compared to anatase and rutile. Appl. Catal. B 200, 647–658.

Uekert, T., Kasap, H., Reisner, E., 2019. Photoreforming of Nonrecyclable Plastic Waste over a Carbon Nitride/Nickel Phosphide Catalyst. J. Am. Chem. Soc. 141, 15201–15210.

Uekert, T., Kuehnel, M. F., Wakerley, D. W, Reisner, E., 2018. Plastic and mixed waste as feedstocks for solar-driven H$_2$ production. Energy Environ. Sci. 11, 2853–2857.

Vequizo, J. J. M., Matsunaga, H., Ishiku, T., Kamimura, S., Ohno, T., Yamakata, A., 2017. Trapping-induced enhancement of photocatalytic activity on brookite TiO$_2$ powders: comparison with anatase and rutile TiO$_2$ Powders. ACS Catal. 7, 2644–2651.

Xu, M., Gao, Y., Moreno, E. M., Kunst, M., Muhler, M., Wang, Y., Idriss, H., Wöll, C., 2011. Photocatalytic activity of bulk TiO$_2$ anatase and rutile single crystals using infrared absorption spectroscopy. Phys. Rev. Lett. 106, 138302.

Yurdakal, S., Palmisano, G., Loddo, V., Augugliaro, V., Palmisano, L., 2008. Nanostructured rutile TiO$_2$ for selective photocatalytic oxidation of aromatic alcohols to aldehydes in water. J. Am. Chem. Soc. 130, 1568–1569.

Zeng, Q., Chang, S., Beyhaqi, A., Lian, S., Xu, H., Xie, J., Guo, F., Wang, M., Hu, C., 2020. Efficient solar hydrogen production coupled with organics degradation by a hybrid tandem photocatalytic fuel cell using a silicon-doped TiO$_2$ nanorod array with enhanced electronic properties. J. Hazard. Mater. 394, 121425.

Zerjav, G., Zizek, K., Zavasnik, J., Pintar, A., 2022. Brookite vs. rutile vs. anatase: What's behind their various photocatalytic activities? J. Environ. Chem. Eng. 10, 107722.

Zhang, J., Yan, S., Fu, L., Wang, F., Yuan, M., Luo, G., Xu, Q., Wang, X., Li, C., 2011. Photocatalytic degradation of rhodamine B on anatase, rutile, and brookite TiO$_2$. Chinese J. Catal. 32, 983–991.

Zhang, L., Menendez-Flores, V. M., Murakami, N., Ohno, T., 2012. Improvement of photocatalytic activity of brookite titanium dioxide nanorods by surface modification using chemical etching. Appl. Surf. Sci. 258, 5803–5809.

Zhao, B., Chen, F., Huang, Q., Zhang, J., 2009. Brookite TiO$_2$ nanoflowers. Chem. Comn. 34, 5115–5117.

Zhou, J., Tian, G., Chen, Y., Wang, J-Q., Cao, X., Shi, Y., Pan, K., Fu, H., 2013. Synthesis of hierarchical TiO$_2$ nanoflower with anatase-rutile heterojunction as Ag support for efficient visible-light photocatalytic activity. Dalton Transactions 42, 11242–11251.